\documentclass[10pt]{article}

\usepackage{amsmath}
\usepackage{amssymb}

\usepackage{graphicx}

\usepackage{cite}

\usepackage{color} 

\usepackage[labelfont=bf,labelsep=period,justification=raggedright]{caption}

\makeatletter
\renewcommand{\@biblabel}[1]{\quad#1.}
\makeatother

\date{}

\begin{document}

\begin{flushleft}
{\Large
\textbf{Are Opinions Based on Science: Modelling Social Response to Scientific Facts}
}
\\
Gerardo I{\~n}iguez$^{1,\ast}$, 
Julia Tag{\"u}e{\~n}a-Mart{\'\i}nez$^{2}$, 
Kimmo K. Kaski$^{1}$,
Rafael A. Barrio$^{3}$
\\
\bf{1} Department of Biomedical Engineering and Computational Science, Aalto University School of Science, P.O. Box 12200, FI-00076 AALTO, Finland
\\
\bf{2} Centro de Investigaci\'{o}n en Energ\'{\i}a, Universidad Nacional Aut\'{o}noma de M\'{e}xico, 62580 Temixco, Morelos, Mexico
\\
\bf{3} Instituto de F\'{\i}sica, Universidad Nacional Aut\'{o}noma de M\'{e}xico, Ciudad Universitaria, 04510 M\'{e}xico Distrito Federal, Mexico
\\
$\ast$ E-mail: gerardo.iniguez@aalto.fi
\end{flushleft}

\section*{Abstract}

As scientists we like to think that modern societies and their members base their views, opinions and behaviour on scientific facts. This is not necessarily the case, even though we are all (over-) exposed to information flow through various channels of media, i.e. newspapers, television, radio, internet, and web. It is thought that this is mainly due to the conflicting information on the mass media and to the individual attitude (formed by cultural, educational and environmental factors), that is, one external factor and another personal factor. In this paper we will investigate the dynamical development of opinion in a small population of agents by means of a computational model of opinion formation in a co-evolving network of socially linked agents. The personal and external factors are taken into account by assigning an individual attitude parameter to each agent, and by subjecting all to an external but homogeneous field to simulate the effect of the media. We then adjust the field strength in the model by using actual data on scientific perception surveys carried out in two different populations, which allow us to compare two different societies. We interpret the model findings with the aid of simple mean field calculations. Our results suggest that scientifically sound concepts are more difficult to acquire than concepts not validated by science, since opposing individuals organize themselves in close communities that prevent opinion consensus.


\section*{Introduction}

Human societies are examples of dynamical complex systems where the complexity stems, on one hand, from the social interactions between individuals of the society, and on the other from societies' organizing structure, making them challenging to study empirically. In these systems the network approach and mathematical modelling have turned out to be crucial in gaining understanding of their structure and dynamical behaviour~\cite{newman2006,caldarelli2007,dorogovtsev2010}. This has been achieved by using such techniques as the analysis of large data sets, collected in questionnaires or obtained by automatically recorded digital footprints of our actions. One of the interesting phenomena in a society - describable as social networks of individuals and links between them - is the process of opinion formation among people, where it is assumed that people focus on a single question answerable simply by yes or no. Likewise one can consider a rather similar situation of spreading scientific information among people, such that during the process some people adopt it as truth and some others reject it. To address these issues there has recently been increasing interest in modelling opinion formation, with many examples in the literature~\cite{castellano2009,sobkowicz2009,weidlich1991,holley1975,sznajd2000,deffuant2000}. Here we extend this approach to study the response of people to scientific facts in co-evolving social networks~\cite{gross2009,perc2010,holme2006,nardini2008,vazquez2008} and under the influence of mass media~\cite{holyst2001sim,kuperman2002srm,sobkowicz2009sos}.

In a recent paper~\cite{iniguez2009} we have described a new model for opinion formation, based on co-evolution dynamics with a single continuously-changing state variable (measuring the degree of agreement with a single question posed) and with link rewiring, i.e. changing the structure of the network. These two mechanisms take place with different time scales such that the former stands for a fast {\it transaction} time scale and the latter for a slow {\it generation} time scale. The rewiring mechanism is based on simple rules adopting the social network dynamics proposed by Kossinets and Watts~\cite{kossinets2006}, that is, a drive for the closure of network triangles and similarity of individuals' opinions. Our model was designed to capture some of the most essential features of opinion formation, namely, dyadic interactions corresponding to one-to-one discussions, an attitude parameter describing the personal reaction of an individual to the overall opinion, and an external influence on the whole network of individuals mimicking the effect of media. In our previous studies~\cite{iniguez2009,iniguez2011cpc} we ignored the role of an external field to simplify the treatment of the model. In reality, however, it is difficult to envisage a situation in which the external field is zero, since there is always information available to all individuals that can affect their decision-making processes.

A good example of this situation is how much scientific and technological knowledge determines the public engagement with polemic issues that affect society. Although we live in societies deeply intertwined with modern science and various emerging technologies, most of the world's population is scientifically illiterate, as can be inferred from examining children of school age~\cite{oecd2012}. In contrast one would expect a literate citizen to be able to evaluate the quality of scientific information, take rational positions on polemic decisions, and make a distinction between science and pseudo-science. A very important goal of science literacy and its public understanding is precisely to erase scientific mistakes and superstitions~\cite{miller2004}.

In order to understand the public perception of science and technology different models have been built, but there is still some controversy between the two main approaches~\cite{nisbet2007}. First, the science literacy model assumes that knowledge increases the public acceptance of scientific research and allows scientists to work with more freedom. As a consequence a knowledgeable public would reach consensus with experts quite easily. The second is a ``contextualist" approach to the public understanding of science, where social values and attitudes of the public are also very important. This is the so-called ethnographic case study method, where alternative forms of knowledge interact with the actions of experts. Quantitative surveys are often related to the science literacy model and generalized attitudes, while the ethnographic approach emphasizes local knowledge. This conflict has led to the dismissal of either surveys or ecological data, although the two models can complement each other. In~\cite{nisbet2007} a survey-based approach was proposed to integrate multiple influences and bridge this gap.

In a recent study~\cite{allum2008} a comparison between roughly 200 surveys from 40 countries was made, finding a correlation between knowledge of scientific facts and a favourable attitude of the public towards science. However, it also turned out that knowledge itself seems to have just a small influence on the opinion. Instead moral values, religious beliefs, trust and cultural trends of communities were found to be significant factors for building the final attitude of an individual. This study concluded that more research needs to be done on the relationship between knowledge and attitude towards science.

It is commonly considered desirable that the development of science in a country should be linked to a policy of science communication,  leading to an increase of the public understanding of science. The term ``science communication" is used here as a technical term to describe the multidisciplinary field of science dissemination. In a recent book on the matter~\cite{bennet2011} the editors report  several stories of success in science communication, covering a wide range of topics in which this link is associated with success. They also point out that the scientific community is increasingly aware of the importance of such a link. However, the activities related to communicating science to a broad audience are not based on quantitative models, and the main source of data is still heavily based on surveys~\cite{schiele2012}. In present day surveys very little attention is put into extracting dynamical changes in the general knowledge of the public, which are extremely important to testing the success of different policies.

In general, there is a great interest in how people react to new technologies, how concerned citizens are about their risks, and how aware they are of their benefits. For instance, in the case of nanotechnology a recent survey in USA~\cite{cacciatore2011} found the opinion of the public to be quite favourable. This study shows that if there is an effort to give simple and accurate information the public will react in a positive way, even enabling them to associate nanotechnology with other fields such as medicine, machinery or computing, and thus tending to have positive opinions and a perception of low risk.

As a summary of the research performed towards the public understanding of science, we list the main factors known to influence social opinion:
\begin{itemize}
\item {\em Knowledge}. In \cite{allum2008} a weak correlation between knowledge of scientific facts and positive attitude towards science was found, but also that it can become negative in case of certain technologies, such as in the human embryo research. These results show that people who are more scientifically literate have more positive attitudes towards science in general, but can be against a specific application or scientific research.
\item {\em Science Communication}. In relation to this factor the quality and relevance of science communication information could move opinion from negative to positive, as in the case of nanotechnology~\cite{cacciatore2011}.
\item {\em Culture}. When different countries are compared social, political, and cultural disparities between nations affect the results in fundamental ways. Local ideology is yet another factor. 
\end{itemize}
All these factors combined affect the personal opinion. There is also a relation with regional socio-economic conditions: in the developing countries science is often idealized, while in post-industrial societies there is not such a positive cultural stereotype~\cite{allum2008}. 

In addition to the science perception models discussed above, a remark must be made on cognitive models. Many of these~\cite{nisbet2007,cacciatore2011} are based on the cognitive miser model, where citizens are supposed to collect only as much information about the topic as they think is necessary to reach an opinion~\cite{fiske2010}. Most of the citizens avoid deep information but rely more on their ideology together with information from mass media. The news, reported in different media, is the main component for the opinion formation process. The cognitive miser model of social thinking reflects the importance of motivation and emotions and the reluctance to do much extra thinking. Besides this superficial associative reasoning there is a deeper cognitive process, which can be considered true reasoning. The mind has two different modes of processing, one automatic and another conscious, which in a way work together. Some situations require the individual to overcome unprecedented obstacles, as happens in scientific research, and the conscious mind takes the challenge. This explains the attitude of experts towards certain problems. As for the two cognitive processing modes~\cite{baumeister2011}, the average citizen is usually in the cognitive miser state most of the time.

Considering all these factors and cognitive models, the main motivation of the present work is to use the knowledge gathered by scientists in different fields, including social scientists, to construct a mathematical model that could lead to the quantification of the process of public understanding of science. With this information in mind we have extended our opinion-formation/co-evolution model to simulate the effects of external information on perception spreading among individuals in a social network. The key ingredients are the following: Each citizen has a personal opinion that is related not only to his or her scientific education but also to the cultural background, and that is likely to change through personal one-to-one interactions. There is an external influence coming from media and science communicators represented by a field. Finally, as the long-range order and interactions are also important and the perception of the general information is considered personal, they are represented by an attitude parameter that corresponds to the reaction of an individual to what friends, relatives, and colleagues think.

This paper is organized such that first we describe the model we have used to simulate the effect of external bits of information on the opinion formation in a social network of individuals. Then we study the model's behaviour via numerical simulations and analytical approximations. After that we adjust the field strength by using data from two survey studies realized as questionnaires. Finally we discuss our results and present concluding remarks.

\section*{Results}

\subsection*{Description of the model}

As our model to simulate the effect of external information coming from media on the formation of perception of an issue in a social network of interacting individuals, we take the co-evolving opinion formation model first studied in~\cite{iniguez2009}. This model is used since we consider it to be quite generic in nature, and straightforwardly extendable to include the influence of media by introducing an external field that affects all individuals of the society. Thus the equation for the state variable of each agent $i$ can be written as,
\begin{equation}
\label{eq:1}
\frac{dx_i}{dt} = \frac{\partial x_i}{\partial t} + \sum_j \hat{O}(x_i, x_j, g) A_{ij},
\end{equation}
where the operator $\hat{O}$ stands for the rewiring operations performed on the adjacency matrix $A_{ij}$ of the social network describing the pairwise interactions between agents $i$ and $j$. These changes occur after $g$ opinion transactions have taken place, defining a slow time scale $T = gdt$ for the network evolution. Each agent is characterized by a state variable $x_i$ representing the instantaneous inclination of the personal perception or opinion on a posed question or issue, and it is bounded between -1 (total disagreement) and 1 (total agreement). The fast dynamics of each state variable is described as follows,
\begin{equation}
\label{eq:2}
\frac{\partial{x_i}}{\partial{t}} = \alpha_i f_l(\lbrace x_j \rbrace_l) + x_i f_s(\lbrace x_j \rbrace_s) + h_i,
\end{equation}
where the first term on the right-hand side represents the reaction of agent $i$ to the overall opinion of all distant individuals $\lbrace x_j \rbrace_l$, modulated by the agent's own attitude ($\alpha_i$) towards the overall or public opinion $f_l$. Hence we can write,
\begin{equation}
\label{eq:3}
f_l = \sum_{\ell = 2}^{\ell_{max}} \frac{1}{\ell} \sum_{j \in m_{\ell}(i)} x_j,
\end{equation}
where $m_{\ell}(i)$ means the set of nodes that are $\ell$ steps away from node $i$, and $\ell_{max}$ is the number of steps needed to reach its most distant neighbours. Note that this quantity is different for each agent, since it depends on the local topology of the network. The second term on the right-hand side of Eq.~(\ref{eq:2}) represents discussions between pairs of linked agents, such that,
 \begin{equation}
\label{eq:4}
f_s = \text{sgn}(x_i) \sum_{j \in m_1(i)} x_j.
\end{equation}
The resulting absolute value $|x_i|$ in Eq.~(\ref{eq:2}) assures that agents with the same sign of opinion reinforce their convictions after a discussion, while agents with opposite sign become less convinced of their position. These are considered as reasonable assumptions based on simple social interactions. The last term $h_i$ on the right-hand side of Eq.~(\ref{eq:2}) is an external field representing the personal bias towards either opinion (-1 or +1), due to e.g. mass media (newspapers, TV, radio), and will be discussed further at the end of this section. 

Note that $|x_i|$ in Eq.~(\ref{eq:2}) could eventually become larger than one, which we choose to interpret as a state of {\it total conviction}. Thus the $x$ variables of totally convinced agents are set to the corresponding extreme value attained ($\pm1$) and their dynamics stopped. These agents cannot modify their state in subsequent times, but they are still linked to the network and taken into account in the dynamical evolution of the undecided agents. In this sense such limit values can be interpreted as final states of irrevocable decision. 

The rewiring scheme involves cutting certain links and creating new ones, as explained in detail in~\cite{iniguez2009}. At the time of cutting an agent $i$ preferentially breaks its link with agent $j$ if there is large disagreement, as quantified by $p_{ij} = |x_i - x_j|/2$. Explicitly, its neighbours are chosen in decreasing order of the opinion difference $p_{ij}$ for $p_{ij} > 0$. Then the agent creates the same number of new links based on either of two link-formation mechanisms known as \textit{triadic} and \textit{focal closure}~\cite{kossinets2006}. In the former it closes a triangle with the ``friend of a friend'' if the new link can help the agent in reaching a state of total conviction (as measured by $q_{ij} = |x_i + x_j|/2$), while in the latter it creates a link with a further neighbour provided their opinions are similar (according to $r_{ij} = |x_i - x_j|/2$). Like in the cutting procedure, new links are created in decreasing order of the opinion similarities $q_{ij}$ and $r_{ij}$ for $q_{ij}$, $r_{ij} > 0$.

In the rewiring scheme used here, agents can create links using these two mechanisms at will. In order to control the proportion between focal and triadic closure events we introduce a quantity $y \in [0, 1]$, which can be used as a stochastic decision parameter for each event. In other words, before creating new links an agent chooses a random number $\xi \in [0, 1]$ from a uniform distribution, and then uses the focal closure mechanism if $\xi < y$ or otherwise the triadic closure mechanism. Note that focal closure was not considered in~\cite{iniguez2009,iniguez2011cpc} for the sake of clarity, since in the absence of an external field opinion formation is practically dominated by close interactions with neighbouring agents. This might not be the case when the important feature is the external information given to all agents in the network, thus we will investigate the role of $y$ here.

In order to apply this model to our present situation we need to reinterpret the meaning of $x$ and give a reasonable form for the external field representing the information given to the social group. The field has to break the symmetry of the model between positive and negative $x$, since now the state variable represents the degree of the individual agent's perception of scientific knowledge. We interpret $x = 1$ as \textit{expert} knowledge, only attained by learned individuals in the field of science of the question or issue. On the other hand \textit{ignorance} is represented by a very small magnitude $|x| \sim 0$, whereas $x < 0$ describes people inclined to disagree with the sound scientific information or agree with unsound concepts, e.g. superstition. Total opposition to scientific truth corresponds to $x = -1$ and we shall refer to such agents as \textit{fundamentalists} from now on, since their attitude against new scientific evidence only reinforces their beliefs and instead of learning, they are stubbornly defending their position against.

As the external scientific information is represented by the field $h_i$ in Eq.~(\ref{eq:2}), we assume that $h_i > 0$ represents a drive towards scientific truth, while $h_i < 0$ describes a drive towards fallacy. The way this information is perceived by a particular agent should then be asymmetric, in the sense that it should help agents to learn and agree with the scientific truth. The simplest way to appropriately break the symmetry is by the linear expression,
\begin{equation}
\label{eq:5}
h_i = h (1 - x_i),
\end{equation}
where $h$ is a constant strength representing the raw external information given to everybody in the network, and the term in brackets reflects an instantaneous personal reaction to the field. Observe that agents with positive $x$ are less affected by this term, while the ignorant and fundamentalist agents are very much in conflict with the field, either positive or negative. This mimics the fact that superstitious people are in general more prepared to believe anything without proof, and change their position with ease. We have also tested quadratic and cubic expressions for the field and the qualitative behaviour of the model remains unchanged, although its analytic treatment becomes unnecessarily complicated. Thus, we have opted for the simpler term in Eq.~(\ref{eq:5}).

It should be noted that the effect of the attitude parameter $\alpha_i$ is not directly affecting the response to the field, but it becomes extremely important when the third term in Eq.~(\ref{eq:2}) compensates the second term, producing an exponential approach to the opinion limits dictated by the first term. In what follows we exhibit numerical results for this model.

\subsection*{Effect of external field or ``mass media''}

We have performed extensive numerical calculations using the model described above. We first adjusted its parameters to produce the results already presented in~\cite{iniguez2009, iniguez2011cpc}, where opinions and network evolve in the absence of an external field ($h = 0$). We initialized the system to a random network configuration of $N = 200$ nodes and average degree $\left< k_0 \right> = 5$, and kept it so for all calculations. We have tried networks of various sizes, and found that the effects of size scale exactly in the same way as the original model without field~\cite{iniguez2009}. Accordingly, we fixed a set of random values for the individual attitudes $\alpha_i$ from a uniform distribution with center $\alpha_c = 0$ and unit half-width, and chose the initial values $x_i(0) = x^0_i$ of the state variables from a Gaussian distribution with zero mean, unit standard deviation and cut-off at $|x| = 1$. With this set of parameter values the network splits in communities sharing the same opinion for $g = 10^3$, where $dt$ is approximated by a time step of size $\Delta t = 10^{-4}$.

\begin{figure}[!ht]
\begin{center}
\includegraphics[width=4in]{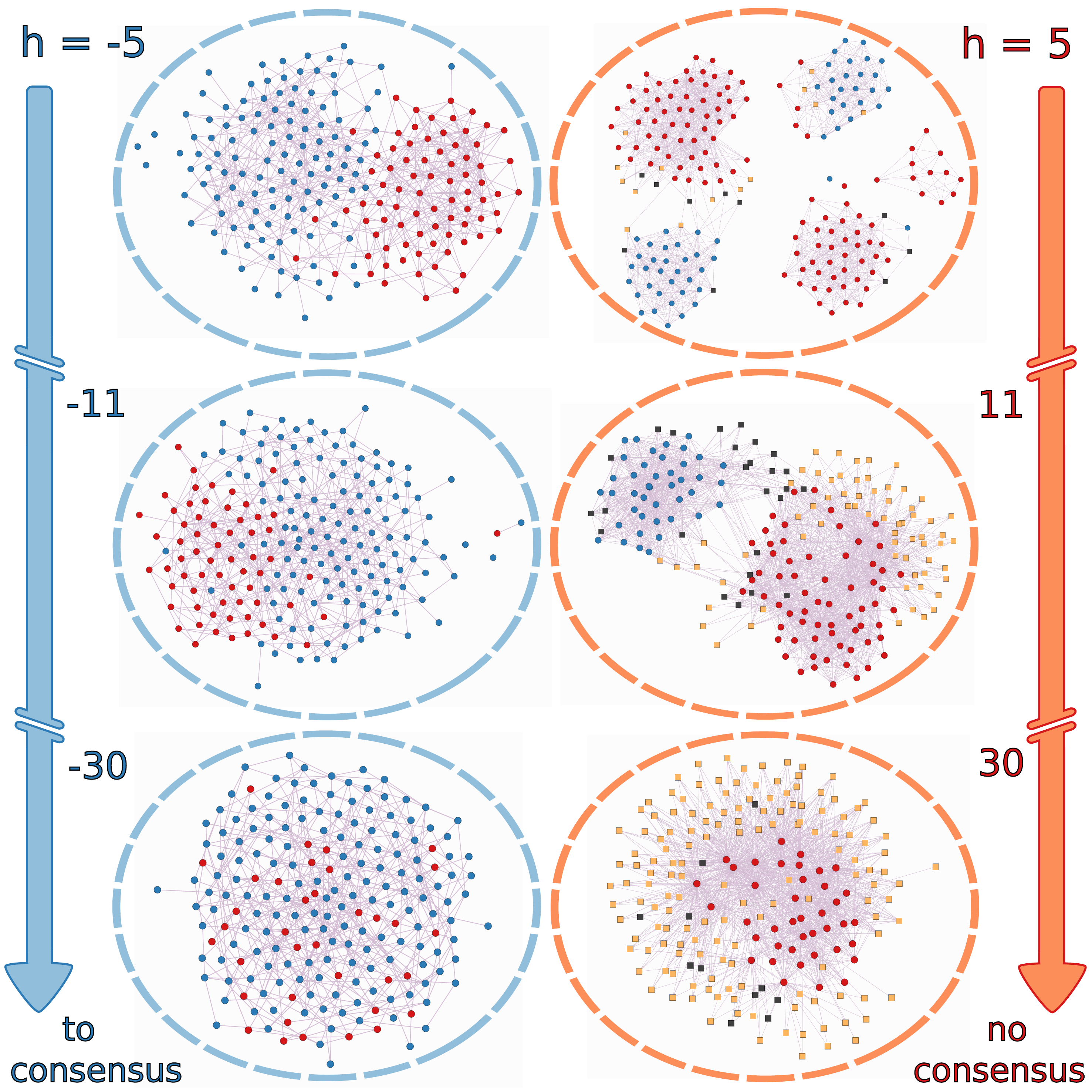}
\end{center}
\caption{
{\bf Final state of the network for $y = 0$.} Asymptotically stationary state of the dynamics for chosen initial conditions and parameters, as described in the text. Decided agents are represented by red ($x_i = 1$) or blue ($x_i = -1$) circles, and undecided agents by yellow ($0 < x_i < 1$) or black ($-1 < x_i < 0$) squares. The right (left) column corresponds to positive (negative) field $h$ of increasing magnitude. All calculations were done with $y = 0$, i.e. rewiring with triadic closure mechanism only. The visualisation shows the basic asymmetry of the model: for stronger negative field the community structure is quickly lost and there is an asymptotic growth towards negative consensus, while for growing positive field fundamentalists linger in well-connected communities and ignorants slow down the drive towards positive consensus.
}
\label{fig:1}
\end{figure}

In order to eliminate the effect of randomness in the calculations we start by setting $y = 0$, i.e. the rewiring takes place only by the triadic closure mechanism. In this way we avoid the need for making averages and thus probe the sole effect of the field $h$ on the system by keeping the initial conditions fixed. In Fig.~\ref{fig:1} we show the himmeli~\cite{himmeli2012} visualisation of the asymptotically stationary final state of the network after simulating $10^5$ time steps of the dynamics. The right and left columns correspond to positive and negative field respectively and for the parameter values described above, as the magnitude of the field $h$ is increased.

These results show interesting effects of the external field on the configuration of the network. First of all, the ratio between the number of experts ($x = 1$) and fundamentalists ($x = -1$) grows for positive field and diminishes for negative field, as expected, but their distribution in the system is not symmetric. This means that for growing positive field quite a few fundamentalists linger in several groups, and all eventually join in a single community that is much more interconnected than the communities of experts. On the other hand, when a negative field grows in magnitude, i.e. an increasing effort to convey non-scientific information to the public, the expert agents opposing the fallacy are dispersed in the network and do not form a community. In terms of the public perception of a concept promoted by raw external information $h$, this implies that scientifically sound concepts (associated with $h > 0$) require a larger field magnitude to create opinion consensus in the network than concepts not validated by science (i.e. $h < 0$), since the fundamentalists found for $h > 0$ organize themselves in close communities that prevent consensus of opinion.

This surprising effect is in agreement with the behaviour of some real social networks (for example, creationists are well organized in very interconnected communities~\cite{park2001}). Furthermore, Fig.~\ref{fig:1} shows that for growing negative field the agents attain a stationary state before having time for rewiring, so the network remains as random as initially but with all of its agents decided already. This situation is asymmetric with respect to the sign of $h$, since community structure is preserved under a growing positive field, and even at high values of $h > 0$ the network's connectivity is larger than is the case for its random counterpart.

Indeed, for progressively stronger positive field the number of undecided agents grows steadily, while for negative field there are zero agents that ``do not know''. These undecided agents (some of them fundamentalists for a weaker positive field) are forced by the large magnitude of $h$ to have the expert opinion but continually resist to do so. Such undecided agents have small $|x|$ values (i.e. behaving as ignorants), and go from being distributed at the edge of the larger cluster of experts to connect the group of fundamentalists with the rest of the network. In this sense the ignorants act as bridges between the tight fundamentalist community and the well-informed people.

\begin{figure}[!ht]
\begin{center}
\includegraphics[width=4in]{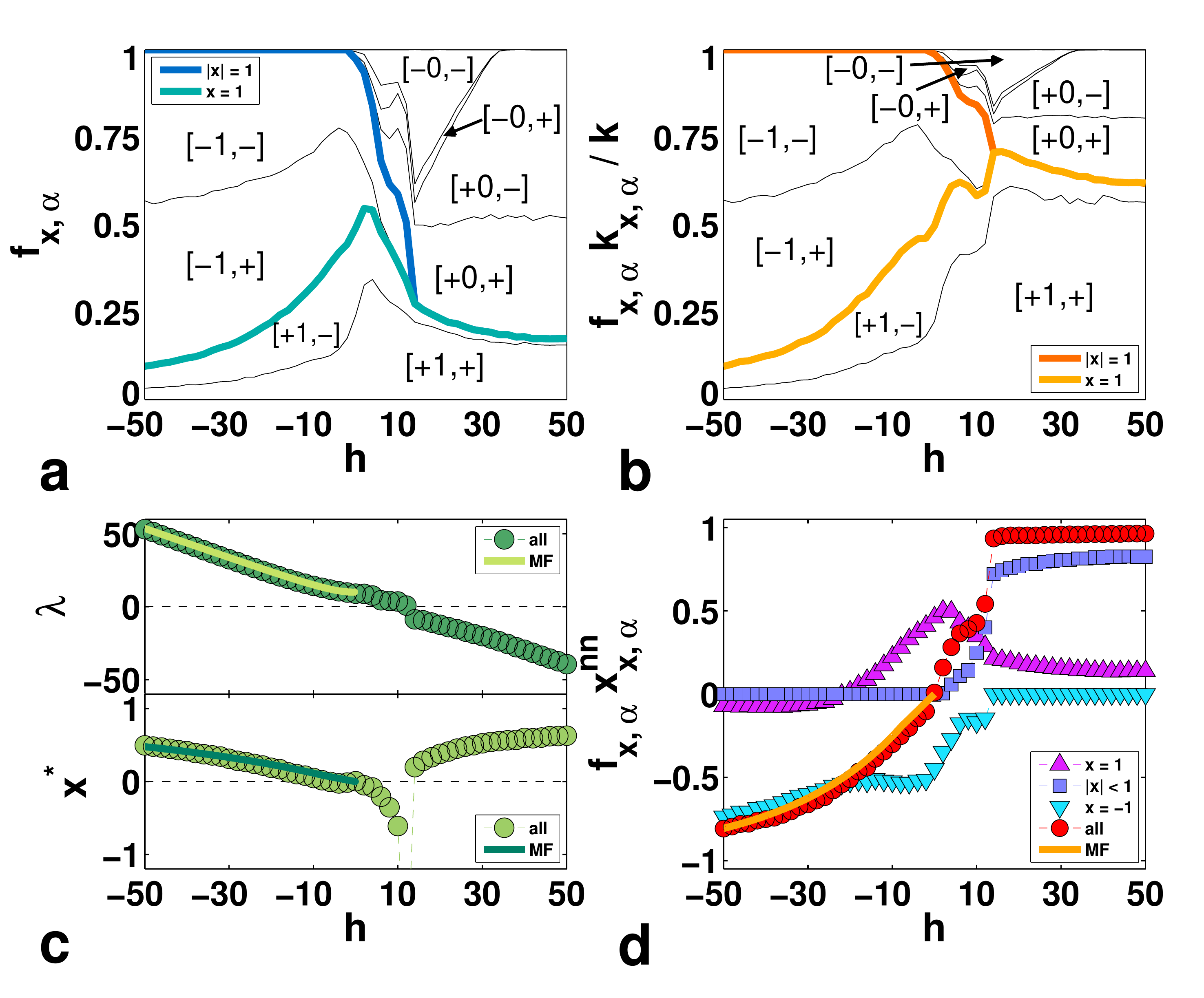}
\end{center}
\caption{
{\bf Effect of external field.} (a) Relative group size as a function of the external field $h$, where each group $[\mathsf{x}, \mathsf{\alpha}]$ is distinguished in terms of both opinion and attitude, as described in the text. Thin lines separate groups and thick lines divide the contributions of experts, ignorants and fundamentalists. (b) Relative group contribution to the average degree of the network as a function of $h$. (c) Fixed point of Eq.~(\ref{eq:2}) (bottom) and associated eigenvalue (top) as a function of the external field. Numerical results are shown as dots, while the corresponding analytical approximations are depicted as lines. (d) Weighted nearest-neighbours' average opinion for agents in group $[\mathsf{x}, \mathsf{\alpha}]$ as a function of $h$, shown as symbols for different opinion groups. The continuous line depicts the analytical approximation of its mean value over all agents. All numerical calculations are averaged over 108 realisations of the dynamics with $y = 0$.
}
\label{fig:2}
\end{figure}

In order to analyse the relationship between opinion and attitude in our model systems, we look at the ensemble averages and classify the agents into groups denoted by $[\mathsf{x}, \mathsf{\alpha}]$. Here the labels take the symbolical values $\mathsf{x} = +1, +0, -0, -1$ for experts, ignorants with positive opinion, ignorants with negative opinion, and fundamentalists, respectively, and $\mathsf{\alpha} = +, -$ for attitude parameter $\alpha_i > \alpha_c$ and $\alpha_i < \alpha_c$. In Fig.~\ref{fig:2}a we plot the fractions $f_{\mathsf{x}, \mathsf{\alpha}}$ of agents in each group $[\mathsf{x}, \mathsf{\alpha}]$, to show the relative group sizes after the normalisation condition $\sum f_{\mathsf{x}, \mathsf{\alpha}} = 1$. Here the thin lines are drawn to distinguish all eight possible groups by value of $\mathsf{x}$ and $\mathsf{\alpha}$, while thick lines are drawn to separate the contributions of experts, ignorants and fundamentalists. It is clear that a negative external field produces roughly symmetrical attitude distributions in groups of decided agents, a steady asymptotic growth to negative consensus, and a total lack of undecided agents. On the other hand, for growing positive field it turns out that experts have mainly positive attitude, all fundamentalists have negative attitude, and there is a large amount of ignorants.

The asymmetry due to the external field is even more evident in the actual topology of the system. In Fig.~\ref{fig:2}b we show the relative contributions of the different groups to the average degree of the network ($k$) as measured by the quantity $f_{\mathsf{x}, \mathsf{\alpha}} k_{\mathsf{x}, \mathsf{\alpha}} / k$, where $k_{\mathsf{x}, \mathsf{\alpha}}$ is the average degree of agents in group $[\mathsf{x}, \mathsf{\alpha}]$. In this case we have a similar normalisation condition, $\sum f_{\mathsf{x}, \mathsf{\alpha}} k_{\mathsf{x}, \mathsf{\alpha}} / k = 1$. The symmetry in attitude sign for negative field is also seen in the left part of Fig.~\ref{fig:2}b, but for positive field the undecided agents (forming the majority of the network, as shown in the right part of Fig.~\ref{fig:2}a) contribute relatively little to the total degree, since most connections belong to expert agents with positive attitude. Moreover, the fundamentalist community found for low $h > 0$ has many connections when compared to its size.

In addition to the simulations, considerable insight into the behaviour of the coupled dynamics of opinion and network structure can be gained with a relatively simple yet analytical mean-field approach, described in detail in section Materials and Methods. It turns out that Eq.~(\ref{eq:2}) has an asymptotic solution $x_i = (x_i^0 - x^*) e^{\lambda t} + x^*$, where the fixed point $x^* = ( h + \alpha_c f_l ) / ( h - f_s )$ and its associated eigenvalue $\lambda = f_s - h$ can be derived explicitly in terms of $h$ and the other parameters of the model. As shown in Fig.~\ref{fig:2}c, increasing the magnitude of negative field results in $\lambda > 0$ and a corresponding repulsive $x^*$ that grows towards $x = 1$, thus giving rise to a diminishing amount of experts and building negative consensus. For increasing positive field the eigenvalue changes sign at a critical value $h_0 \sim 10$, signalling a transition to a state where an attractive fixed point hinders most agents from attaining extreme values of opinion. As $h$ grows $x^*$ moves once again towards $x = 1$, implying positive consensus in the limit of infinite field. Fig.~\ref{fig:2}c shows good agreement between the numerical computation of $x^*$ and $\lambda$ and their analytical approximations.

It is also instructive to look at the effect of $h$ on the weighted nearest-neighbours' average opinion for agents in group $[\mathsf{x}, \mathsf{\alpha}]$, denoted by $f_{\mathsf{x}, \mathsf{\alpha}} x^{\mathsf{nn}}_{\mathsf{x}, \mathsf{\alpha}}$. As seen in Fig.~\ref{fig:2}d, the change of sign in the experts' $x^{\mathsf{nn}}_{\mathsf{x}, \mathsf{\alpha}}$ for increasing magnitude of the negative field implies that they are sparsely distributed in the system, while for $0 < h < h_0$ the fundamentalist group is tightly interconnected and manages to keep a negative nearest-neighbours' average opinion until its disappearance. A phase change in the number of undecided agents (reminiscent of the one discussed in~\cite{iniguez2011pre}, in that case as a function of the time-scales' ratio $g$) also shows in the value of $x^{\mathsf{nn}}_{\mathsf{x}, \mathsf{\alpha}}$ for both groups and the network as a whole. This is signalling a very slow transition to positive consensus as the positive field is increased.

\subsection*{Effect of focal versus triadic closure}

\begin{figure}[!ht]
\begin{center}
\includegraphics[width=4in]{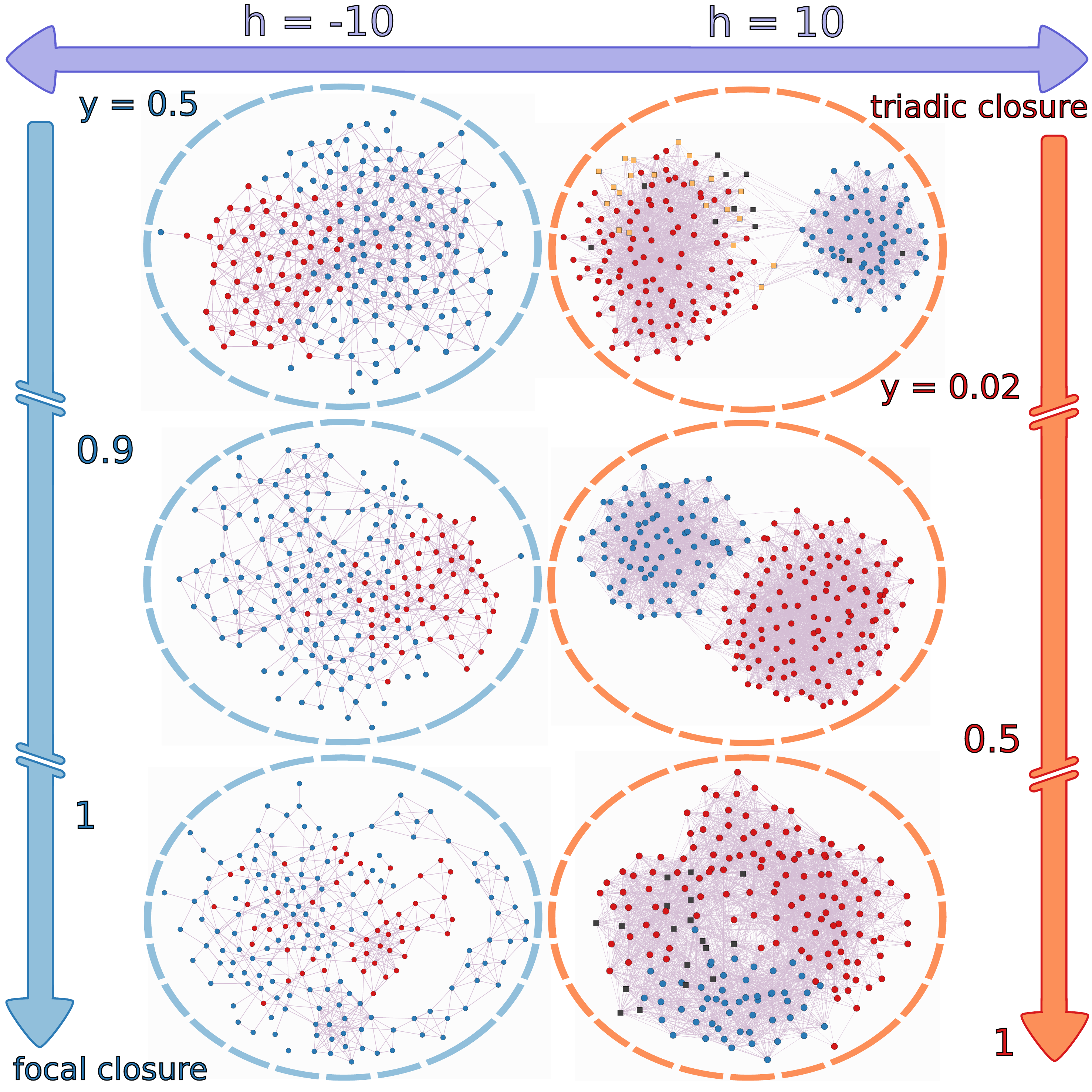}
\end{center}
\caption{
{\bf Final state of the network for $y \neq 0$.} Asymptotically stationary state of the dynamics for the same initial conditions and parameters as in Fig.~\ref{fig:1}. Decided agents are represented by red ($x_i = 1$) or blue ($x_i = -1$) circles, and undecided agents by yellow ($0 < x_i < 1$) or black ($-1 < x_i < 0$) squares. The left and right columns correspond to $h = -10, 10$, respectively, as the parameter $y$ is varied from zero to one. An increasing amount of focal closure events results in a systematic loss of heterogeneous structure in the network, higher degree, and decreasing amount of ignorants.
}
\label{fig:3}
\end{figure}

We now turn to explore the changes introduced by varying the parameter $y$ from zero to one, measuring the relative amount of focal closure instead of triadic closure mechanisms used by agents in the network rewiring process. The effect of $y$ in the final state of the opinion dynamics is illustrated in Fig.~\ref{fig:3} for external field $h = -10, 10$. We used the same set of initial conditions and parameters as in Fig.~\ref{fig:1}, keeping just the inherent randomness associated with non-zero values of $y$. The main difference with the $y = 0$ case is due to the fact that when $y$ grows it is easier for agents to find someone to create a new link with, resulting in a systematic loss of heterogeneous structure. Indeed, for $h \geq 0$ and as $y$ is increased communities of the same opinion quickly merge, the average degree in the network and the number of connections between the two remaining clusters grow, and the number of ignorants decreases. In the case of negative field the network's structure is basically random for most values of $y$, except for $y \sim 1$ where the distribution of connections is more heterogeneous.

There are several qualitative properties of the system that hold for all values of $y$. First, the number of experts for positive field is always smaller than the number of fundamentalists for negative field of the same magnitude. One could infer from this that a true scientifically sound concept is more difficult to acquire than a wrong concept without scientific content. Second, the agents with no neighbours (``outcasts'', so to speak) are usually the first agents to become decided in the dynamics, either experts or fundamentalists, and their number grows with $y$. Third, the communities of fundamentalists for positive field are more tightly connected than the corresponding groups of experts for negative field of the same magnitude, or in other words, disagreeing individuals are more sparsely distributed in the network for $h < 0$ than for $h > 0$. Finally, the undecided agents for positive field surround the largest cluster of experts and serve as bridges to the fundamentalist community.

\begin{figure}[!ht]
\begin{center}
\includegraphics[width=4in]{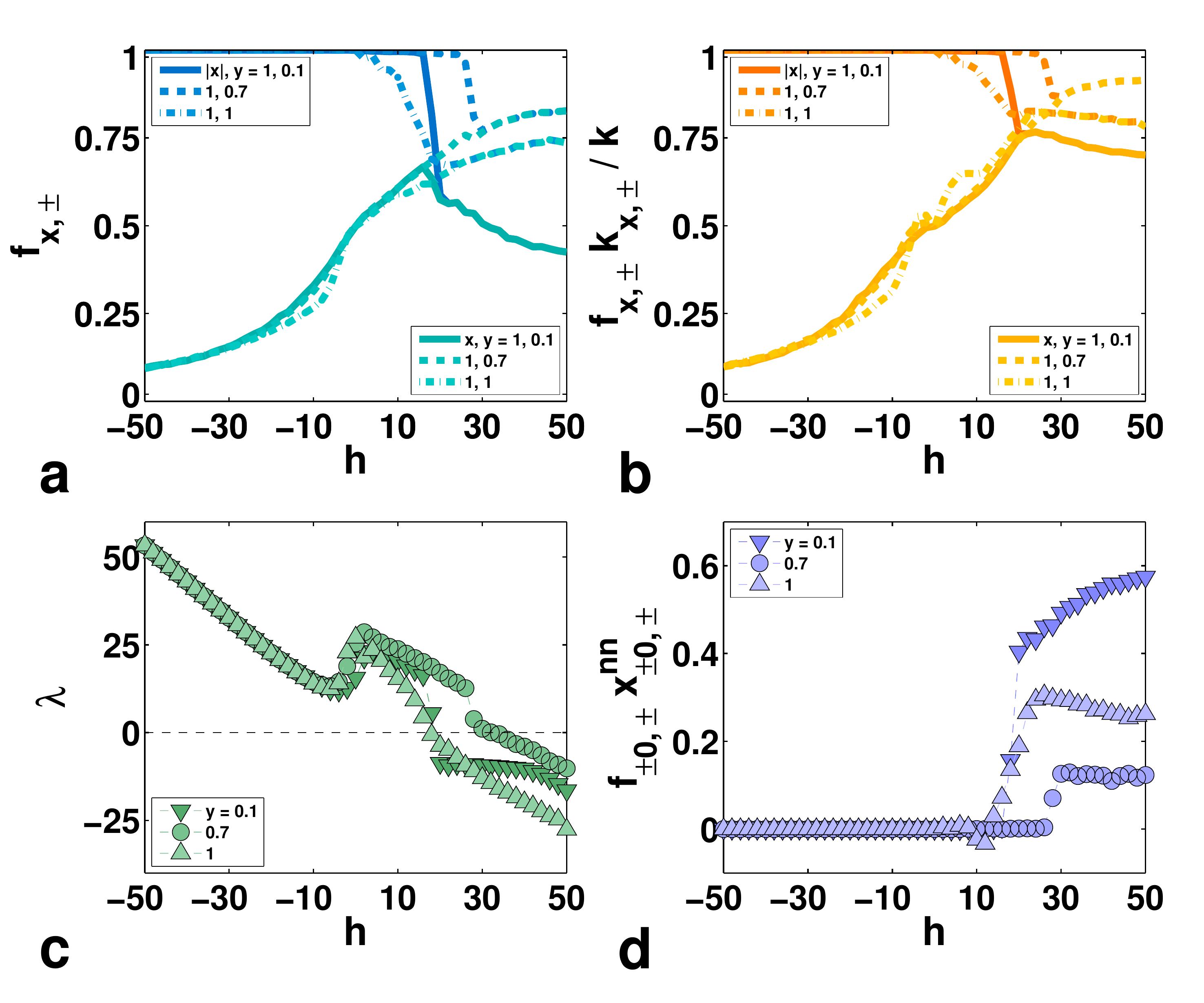}
\end{center}
\caption{
{\bf Focal versus triadic closure.} Relative group size for experts, ignorants and fundamentalists (a), relative group contribution to the average degree of the network (b), eigenvalue associated with Eq.~(\ref{eq:2}) (c), and weighted nearest-neighbours' average opinion for undecided agents (d), all plotted as functions of the external field $h$ and for rewiring parameter values $y = 0.1, 0.7, 1$. All calculations are averaged over 108 realisations of the dynamics. Non-zero values of the rewiring parameter retain the qualitative picture of the $y = 0$ case shown in Fig.~\ref{fig:2}, along with a displacement of the critical point $h_0$ and non-trivial behaviour in the region $0 < h < h_0$.
}
\label{fig:4}
\end{figure}

To further validate these results and take into account the relationship between opinion and attitude, we again perform averages over the relevant random initial conditions and parameters. In Fig.~\ref{fig:4}a we plot the effect of $y$ on the relative contributions of experts, ignorants and fundamentalists, corresponding to the thick lines in Fig.~\ref{fig:2}a. Here we see that the number of undecided individuals gets minimized around $y \sim 0.7$. This particular ratio between focal and triadic closure mechanisms optimizes the presence of experts for $h > 0$, although there is a persistent group of ignorants with both negative opinion and attitude constituting roughly $20\%$ of the network. This is reminiscent of a result in the 2005 Eurobarometer~\cite{euro2005}, where the validity of 13 scientific-knowledge statements (such as ``It takes one month for the Earth to go around the Sun'') was asked and the average percentage of wrong answers amounted to $21\%$. 

The minimum in the number of ignorant agents also shows in the relative group contributions to the average degree of the network, as shown in Fig.~\ref{fig:4}b. The fundamentalist group found in the $0 < h < h_0$ region grows with $y$, has mainly agents with negative attitude and a relatively high amount of connections and triangles. Quite surprisingly, the average clustering coefficient $c$ of the network (not shown here) does not diminish when $y \neq 0$, but increases. This happens in spite of the fact that focal closure is not an explicit mechanism for the formation of triangles.

Finally, in Fig.~\ref{fig:4}c and d we can see the effect of $y \neq 0$ on the eigenvalue $\lambda$ associated with Eq.~(\ref{eq:2}) and the weighted nearest-neighbours' average opinion of undecided agents $f_{\pm0, \pm} x^{\mathsf{nn}}_{\pm0, \pm}$, as opposed to their values for $y = 0$ shown in Fig.~\ref{fig:2}c and d. In Fig.~\ref{fig:4}c we note that the critical point $h_0$ (signalled by a change of sign in $\lambda$) moves with $y$, due to a non-trivial increase in the eigenvalue for the region $0 < h < h_0$, and attains a maximum value of $\sim 30$ for $y \sim 0.7$. In Fig.~\ref{fig:4}d we observe that the position of the phase change in the number of undecided individuals (occurring at zero field for $y = 0$) is also displaced with increasing $y$, and maximized to a value close but less than $h_0$. Incidentally, this also happens for $y \sim 0.7$ and thus coincides with the minimum in the number of ignorants after the phase change has taken place.

In the next section we will discuss and interpret these results in the light of social behaviour, by adjusting the field strength $h$ with the help of actual data extracted from extensive polls made in Mexico and Europe.

\subsection*{Adjusting field strength with survey data}

Surveys and polls do not exactly reflect public opinion, but at least they provide some measure of the public perception about the subjects under investigation. The reasons for the inaccuracies are many, not the least how the survey was conducted including what was asked and how the questionnaire was designed. Most surveys could be regarded as snapshots of society without dynamical information. The survey carried out by the National Science Foundation~\cite{roper2012} is an exception, where the data has been integrated from 1979 to 2001. Furthermore, it is difficult to follow the time development of opinion, as there has been a shift from measuring mostly literacy to matters concerning science and society~\cite{bauer2007}, i.e. from concentrating on knowledge questions to asking about attitudes and trust.

It is quite common that the results of the survey are presented by giving percentages of the responses of the population to a few answering-options in the questionnaire, without further details on the topology of the underlying social network and of possible relations between its individuals. This lack of detail about social structure is unfortunate, since (as evident from above) our model could give very rich information about how society is organized and functioning. Irrespective of these shortcomings the surveys are considered useful, since we can adjust the field strength in our model to give the same percentages of a population's responses as in the actual survey data, and then compare the resulting network topologies that would correspond to surveys of different populations.

Here we will use two different surveys: the well-known Eurobarometer (EU)~\cite{euro2010} and a Mexican survey (Mx)~\cite{mex2010}. The Mexican Consejo Nacional de Ciencia y Tecnolog{\'\i}a has carried out surveys on science and technology perception every two years, starting from 2005. The 2009 survey covered the whole country, with subjects aged 18 or more, giving a total of $40 \, 469 \, 253$ citizens (of whom around 55\% are women and 45\% men). Although the Mx survey follows the methodology of the EU survey, the multiple-choice answers are different. In the Mx survey the possible answers to a given statement are: {\it totally agree}, {\it tend to agree}, {\it tend to disagree}, {\it totally disagree} and {\it do not know}, while in the EU survey there is an extra option: {\it neither agree nor disagree}. Thus, in order to perform a comparison between surveys we consider \{{\it tend to agree}, {\it totally agree}\} together, the same for \{{\it tend to disagree}, {\it totally disagree}\}, and exclude the {\it do not know} and {\it neither agree nor disagree} answers, since they depend too much on the protocol. The agreement and disagreement percentages are then normalized by its sum to give the fractions $n_+$ and $n_-$, respectively. 

\begin{figure}[!ht]
\begin{center}
\includegraphics[width=4in]{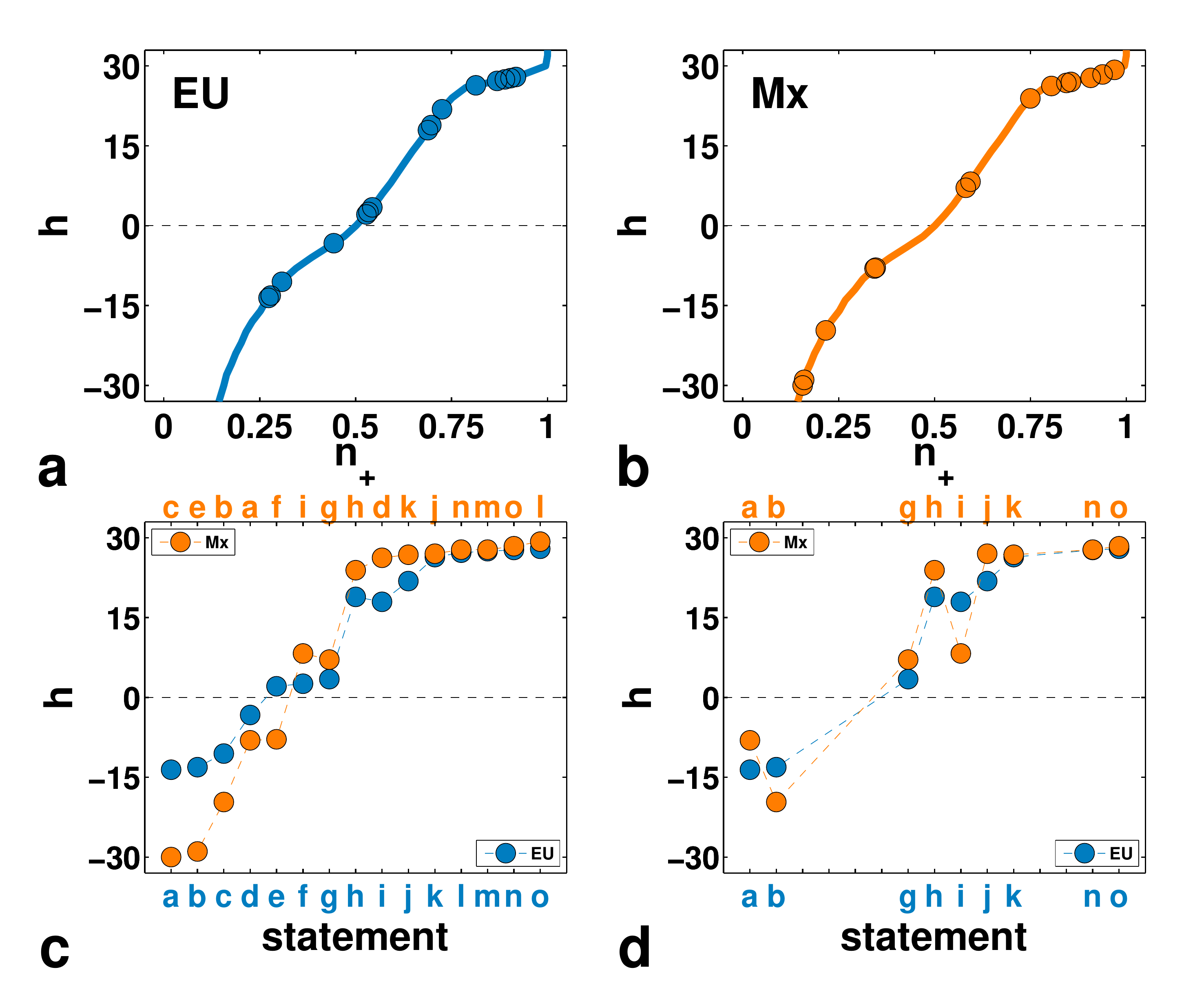}
\end{center}
\caption{
{\bf Adjusting field strength with survey data.} (a) External field strength $h$ as a function of the fraction of agreeing individuals $n_+$ for a rewiring parameter value of $y = 0.7$. Continuous lines show the model simulations for the same set of parameters as used in Fig.~\ref{fig:4}. Symbols correspond to $n_+$ for statements in the EU survey, with adjusted $h$ values so that the agreement fractions between statement and simulation are the same. (b) The same but with symbols corresponding to statements in the Mx survey. (c) Assigned $h$ values for all 15 statements in the EU and Mx surveys, each set in different order according to the statement labels. (d) The same but only for 9 selected statements without cultural content, shown in the same order for both surveys.
}
\label{fig:5}
\end{figure}

The problem in starting a comparison of the model results with survey data is that there is no clear way for assigning an $h$ value to each statement. The main quantitative information at our disposal is the percentage of agreement, and we shall use it in the following way. To assign  an $h$ value to each statement, we select a few statements from the EU survey and use our judgment to order them from blatant fallacies to obvious facts. A list of statements in this order is shown in section Materials and Methods, with their percentages of agreement and disagreement in Table~\ref{tab:1}. The fraction $n_+$ is also calculated from the numerical results and normalized in the same fashion as with the data, which we show as a continuous line in Fig.~\ref{fig:5}a and b. Since this line is mostly a monotonic function, we can give an $h$ value to each statement so that the fractions $n_+$ between survey and model are the same. The same procedure is then followed with the Mx survey, where one finds different values of the field for the same statements. The adjusted values of $h$ for both surveys are shown as dots in Fig.~\ref{fig:5}a and b. Given that our simulations of $n_+$ do not depend very much on the $y$ parameter (describing the proportion between focal and triadic closure), we perform the adjustment procedure by choosing only one value of $y = 0.7$. Notice that the order of the statements in the Mx survey is slightly different, and consequently the values of the field for a given fraction of agreement differ as well.

Of the 15 selected statements, 9 are concerned with purely scientific matters while the other 6 also depend on cultural-related beliefs, although it is indeed difficult to draw a separating line between the two. For instance, the statement ``We depend too much on science and not enough on faith"  shows 38\% agreement in EU and only 16\% agreement in Mx, reflecting cultural differences, while the statement ``Thanks to science and technology, there will be more opportunities for future generations" has 75\% and 88\% agreement in EU and Mx, respectively, with practically the same proportion of disagreement. In Fig.~\ref{fig:5}c we compare the assigned fields of all 15 statements for the two surveys, conserving the order of Table~\ref{tab:1} for the EU survey and changing it for the Mx survey, as indicated by the statement labels. In Fig.~\ref{fig:5}d we keep only the 9 questions without culturally-related content, shown in the same order for both surveys.

First of all it is clear that total agreement or disagreement is never attained, since the dots are bounded by $|h| \sim 30$ (an exceptional case in the Mx survey is ``Smoking can cause lung cancer", where the external field is clearly very strong and produces 98\% of agreement). A peculiar difference is that statements near a 50-50 agreement (i.e. $h \sim 0$) are absent in the Mx survey. This we could interpret as an EU population less affected by propaganda modelled as a field. It is also interesting that polemic statements (for which the response tends to be close to 50-50) are not really distinguishable from statements lacking interest, for which give people give a random answer.

Another point worth mentioning is that although our model predicts an asymmetry between positive and negative field, the results plotted in Fig.~\ref{fig:5}c look more symmetric for the Mx survey than for EU. In terms of the description of our model, this would imply that the Mx population requires a larger external influence to agree to a fallacy. Additionally, we see that for large positive field the behaviour of both populations is very similar, yet for moderately $h > 0$ the EU survey shows a smaller response. Notice that for the non-cultural statements in Fig.~\ref{fig:5}d the behaviour of both populations is also similar, from which we conclude that efforts to inform about matters related to science are comparable once culturally-related beliefs are addressed.

\section*{Discussion}

The value of mathematically modelling a complex system, even in the case of one as intricate as society, is that it allows us to gain understanding of the system's response in situations not quantified in real life before. To this end we have modelled the effect of a controlled external field on the opinion formation process over a co-evolving social network, and we have inferred its consequences in society. As the main feature of our model we conclude that individuals who are stubbornly opposing the spreading of scientific facts have to compensate it by making tightly-connected communities, where they support each other against commonly accepted notions. However, these communities are not completely isolated because ignorant agents serve as bridges connecting different parts of society. This result suggests that scientifically sound concepts are more difficult to acquire than concepts not validated by science, since opposing individuals organize themselves in close communities that prevent opinion consensus.

We have thoroughly analysed the asymmetric response to such field in the dynamics of attaining complete agreement on a scientific fact or correct notion. In the spirit of a mean field approach we could demonstrate analytically that there is a critical field value, beyond which the dynamics is slowed down enormously and a heterogeneous community structure remains. This observation echoes with previous discussions of survey results, where it is seen that very aggressive propaganda does not necessarily result in a proportional immediate increase of agreement.

We have also made an initial attempt at implementing data on scientific perception surveys from two different populations with our model. Even though current surveys do not fully probe the intertwined relationship between social communities and opinion (only measuring an averaged response emanating from the social dynamics), we have used their resulting agreement fractions in science-related statements to adjust the field strength in the model, allowing us to point out differences between the two populations. Particularly, we could infer that a more sceptical society has an asymmetric response in the limits of strong positive and negative external field, and favours divided positions for a weak external field.

A comparison between model results and a real opinion formation process is beyond the scope of the present work, yet constitutes a very interesting and challenging task for the future. To advance in this direction one could analyse the available media content on each particular subject over time and relate it to a quantity like the external field $h$ in our model, as some related studies on this matter suggest~\cite{bauer2005}. Furthermore, current surveys could be enhanced by clarifying the ``do not know'' answers, integrating data over time, and keeping track of the network topology details.

The present study was aimed at investigating the susceptibility of an ethnic group facing such universal notions as science and technology. This problem has occupied the thoughts of many scholars for a long time, and we think it is time to make efforts in quantifying such notions. It is expected that there are cultural differences in answering specific questions, but it is also not surprising to find many similarities due to human cognitive processes and the globalized world we live in. Further insights coming from mathematical modelling, such as those presented here, could be useful in improving current policies to communicate the facts and findings of science in a socially responsible way.  

\section*{Materials and Methods}

\subsection*{Analytical calculations}

We can introduce a series of analytical approximations to shed light over the complex dynamics found in our co-evolving social network model, and help in the understanding of the numerical simulations discussed in the Results section. In order to accomplish this, we analyze the stationary state resulting from the coupling between opinion dynamics and network structure in the spirit of a mean-field approximation.

Let us start by introducing a site-dependent integrating factor,
\begin{equation*}
\mathcal{I}(t) = e^{ \int (h - f_s)d\tau },
\end{equation*}
for the equation of motion $\partial_t x_i + (h - f_s) x_i = \alpha_i f_l + h$, which allows us to write the formal solution,
\begin{equation}
\label{eq:6}
x_i(t) = \frac{1}{\mathcal{I}} \left[ \mathcal{I}^0 x_i^0 + \int_0^t \mathcal{I}(\tau) \left( \alpha_i f_l(\tau) + h \right) d\tau \right],
\end{equation}
where $\mathcal{I}^0 = \mathcal{I}(0)$. In the asymptotically stationary final state of the system, we can substitute the site-dependent short- and long-range interaction terms in Eqs.~(\ref{eq:4}) and~(\ref{eq:3}) by average quantities $f_s$ and $f_l$ that vary slowly in time, so that $\mathcal{I} = \exp \left[ (h - f_s) t \right]$ and Eq.~(\ref{eq:6}) can readily be integrated to give,
\begin{equation}
\label{eq:7}
x_i = (x_i^0 - x^*) e^{\lambda t} + x^*,
\end{equation}
where $x^*$ is the mean fixed point of Eq.~(\ref{eq:2}) and $\lambda$ its associated eigenvalue.

Indeed, a linear stability analysis of Eq.~(\ref{eq:2}) around $x^*$ results in,
\begin{equation}
\label{eq:8}
x^* = \left. x_i \right|_{\partial_t x_i = 0} = \frac{h + \alpha_c f_l}{h - f_s},
\end{equation}
and,
\begin{equation}
\label{eq:9}
\lambda = \left. \frac{\partial( \partial_t x_i )}{\partial x_i} \right|_{x_i = x^*} = f_s - h,
\end{equation}
where the quenched attitude parameter $\alpha_i$ has been substituted by its average $\alpha_c$ and correlations were ignored. These expressions allow us to determine the qualitative behaviour of the system as a function of the external field. Since $f_s$ and $f_l$ are respectively bounded by the degree $k$ and by $N - k - 1$, for $h \ll 0$ the eigenvalue is trivially positive, the repulsive fixed point approaches $x = 1$ and a steady asymptotic growth towards negative consensus is achieved. For $h \gg 0$ this situation is reversed to give a negative eigenvalue, an attractive fixed point approaching $x = 1$, and positive consensus in the limit $h \to \infty$. Since $f_s$ is continuous, there is at least one value $h_0$ such that $f_s(h_0) = h_0$, an implicit definition of the critical point discussed in the Results section. Note that consensus is eventually achieved in both directions, but a fundamental asymmetry arises in the dynamical evolution of the system due to the nature of its fixed point, with only decided agents for negative field and undecided agents for positive field. 

The idea now is to find appropriate expressions for $f_s$ and $f_l$ in terms of $x^*$ and the parameters of the model, then solve for the fixed point and obtain the explicit dependence on the field $h$. To simplify further derivations it is useful to introduce the scaled fixed point $u = b_0 x^*$, where $b_0 = \sqrt{2/\pi} / \text{erf}(1/\sqrt{2})$, and with which the Gaussian distribution of initial opinions can be written as $\rho_0 (x) = (b_0 / 2) \exp(-x^2 / 2)$.

As an intermediate step we estimate the change in the average degree of the network due to the first parallel rewiring of undecided agents. As described with detail in~\cite{iniguez2009}, the simultaneous rewiring of neighbouring nodes can lead to the net creation or deletion of a link. For simplicity we assume $y = 0$ (i.e. only triadic closure events) and $k_0 \ll N/2$, so that the network to be rewired is essentially a Cayley tree with coordination number $k_0$. Since the number of 2nd neighbours of a rewiring agent $i$ is $n_2 = k_0 (k_0 - 1)$ and thus larger than its degree, agent $i$ chooses to cut all of its $k_0$ connections and create $k_0$ links with 2nd neighbours of appropriate opinion. Each rewiring neighbour takes the same decision with probability 1, thus bonds are cut twice leading to an increase in the degree proportional to $k_0$. On the other hand, each newly-rewired neighbour can also choose agent $i$ from its own set of 2nd neighbours to create a link with probability $k_0 / n_2$, thus bonds are made twice leading to a decrease in the degree proportional to $k_0^2 / n_2$. This situation requires the simultaneous rewiring of two undecided agents, then the resulting change in degree is estimated as $k - k_0 = n_0^2 k_0 [1 - 1/(k_0 - 1)]$, with $n_0$ the fraction of undecided agents in the network.

Now, for $h < 0$ and in the presence of a repulsive $x^*$, most agents get decided before the previously discussed first rewiring takes place, so the calculated degree is a good approximation for the final degree of the network. As for $n_0$, it can be estimated by integrating the distribution of initial opinions $\rho_0$ over the interval $[x_g^-, x_g^+]$, where $x_g^{\pm} = x^* \pm (1 \mp x^*) \exp(-\lambda g \Delta t)$ are the positive/negative initial opinions required to reach extreme opinions in exactly $g$ time steps of size $\Delta t$, according to the solution in Eq.~(\ref{eq:7}). The resulting integral can be expanded to first order for $x_g^{\pm} < \sqrt{2}$ and gives $n_0 = b_0 \exp(-\lambda g \Delta t)$. Thus, considering that $f_s \sim 0$ for early times we can write the final degree in the system as,
\begin{equation}
\label{eq:10}
k = k_0 \left[ 1 + \left( 1 - \frac{1}{k_0 - 1} \right) b_0^2 e^{2 g \Delta t h} \right],
\end{equation}
implying an exponential decrease in $k - k_0$ as the magnitude of $h < 0$ increases.

Let us continue and find asymptotically correct expressions for $f_s$ and $f_l$ in terms of the scaled fixed point $u = b_0 x^*$ and the parameters of the model. We denote by $\lbrace s_i \, s_f \rbrace$ the fraction of agents that start with an initial opinion sign $s_i$ and end up in an extreme value of opinion with sign $s_f$. In the presence of the repulsive fixed point $x^*$ found for $h < 0$, the final fraction of experts is $\lbrace ++ \rbrace = \int_{x^*}^1 \rho_0(\tau) d\tau = (1 - u) / 2$, and $\lbrace +- \rbrace = \int_0^{x^*} \rho_0(\tau) d\tau = u/2$ is the fraction of initially positive agents that convert to fundamentalists due to their adverse neighbourhood. On the other hand, $\lbrace -+ \rbrace = 0$ and the fraction of initially negative agents with a favouring neighbourhood is $\lbrace -- \rbrace = 1/2$. Then, the short-range interaction term can be written as $f_s = \lbrace ++ \rbrace f_s^+ + \left( \lbrace -- \rbrace + \lbrace +- \rbrace \right) f_s^-$, where the corresponding contributions by experts and fundamentalists are,
\begin{equation*}
\begin{cases}
f_s^+ &= 2k \left[ \lbrace ++ \rbrace - \lbrace +- \rbrace \right] = k(1 - 2u) \\
f_s^- &= 2k \left[ \lbrace -- \rbrace - \lbrace -+ \rbrace \right] = k
\end{cases},
\end{equation*}
that is,
\begin{equation}
\label{eq:11}
f_s = k \left[ u^2 - u + 1 \right].
\end{equation}
Here we reflect the fact that, in average, the field-favoured negative agents end up as fundamentalists being connected mostly between themselves, while most positive agents are in conflict with the field and cannot find proper neighbourhoods.

We then follow a similar argument for the long-range interaction parameter and write $f_l = \lbrace ++ \rbrace f_l^+ + \left( \lbrace -- \rbrace + \lbrace +- \rbrace \right) f_l^-$, where positive agents contribute with $f_l^+ = a_k (1 - 2u)$ and negative agents with $f_l^- = -a_k$, since $f_l$ does not have a $\text{sgn}(x_i)$ term like $f_s$ and consequently it is asymmetric with respect to opinion sign. The factor $a_k$ measures the average number of further neighbours weighted by their geodesic distance to an arbitrary node $i$, in accordance with Eq.~(\ref{eq:3}), and can be easily calculated for an acyclic network as $a_k = \sum_{\ell = 2}^{\ell_{max}-1} k(k - 1)^{\ell-1} / \ell + R/\ell_{max}$, where $\ell_{max} - 1$ is the largest layer number such that the sum over $\ell$ is less than $N$, and $R$ is the remaining number of nodes in the network, excluding agent $i$. Thus, $f_l$ is estimated as,
\begin{equation}
\label{eq:12}
f_l = -a_k \left[ 2u - u^2 \right].
\end{equation}

Finally, we can use Eq.~(\ref{eq:10}) to write $k$ and $a_k$ explicitly in Eqs.~(\ref{eq:11}) and~(\ref{eq:12}), then insert these in Eq.~(\ref{eq:8}) and solve for $u$. After some algebra we find a cubic equation for the scaled fixed point,
\begin{equation}
\label{eq:13}
ku^3 + ( b_0 \alpha_c a_k - k )u^2 + ( k - h - 2 b_0 \alpha_c a_k )u + b_0 h = 0,
\end{equation}
which can be solved analytically to give $x^* = u / b_0$ as a function of $h$ and the parameters of the model. The result, plotted as a continuous line in the bottom part of Fig.~\ref{fig:2}c, is in close agreement with the numerical simulations. With this solution we can also calculate analytically the eigenvalue $\lambda = k \left[ u^2 - u + 1 \right] - h$, shown in the top of Fig.~\ref{fig:2}c, and the mean value of the nearest-neighbours' average opinion $x^{nn} =\left[ \lbrace ++ \rbrace f_s^+ - \left( \lbrace -- \rbrace + \lbrace +- \rbrace \right) f_s^- \right] / k = u^2 - 2u$, depicted as a continuous line in Fig.~\ref{fig:2}d, again in good agreement with numerical results.

\subsection*{Survey data}

We chose 15 equivalent statements from the EU~\cite{euro2010} and Mx~\cite{mex2010} surveys concerning the public perception of science and technology in society. Statements in both surveys are deemed equivalent apart from minor changes in wording and differences in country names, and according to~\cite{euro2010} they read as follows:
\begin{enumerate}
\item Thanks to scientific and technological advances, the Earth's natural resources will be inexhaustible.
\item Science and technology can sort out any problem.
\item Science and technology cannot really play a role in improving the environment.
\item Science should have no limits to what it is able to investigate.
\item We depend too much on science and not enough on faith.
\item Some number are especially lucky for some people.
\item Scientists should be allowed to experiment on animals like dogs and monkeys if this can help sort out human health problems.
\item The benefits of science are greater than any harmful effects it may have.
\item Because of their knowledge, scientists have a power that makes them dangerous.
\item Science makes our ways of life change too fast.
\item The application of science and new technologies will make people's work more interesting.
\item Compared with research carried out and funded by each Member State, to what extent do you think that collaborative research across Europe and funded by the European Union is in the national interest?
\item Even if it brings no immediate benefits, scientific research which adds to knowledge should be supported by Government.
\item Thanks to science and technology, there will be more opportunities for future generations.
\item A scientific discovery is in itself neither ``good'' nor ``bad'', it is only the way the discovery is used which matters. 
\end{enumerate}

According to our subjective judgment, the previous list is ordered from blatant fallacies to obvious facts. In Table~\ref{tab:1} we show all statements for both the EU and Mx surveys. The percentages of agreement (yes) and disagreement (no) are also shown, such that the remaining percentage corresponds to the sum of the {\it do not know} and {\it neither agree nor disagree} answers for the EU survey, and to the {\it do not know} answer for the Mx survey.

\section*{Acknowledgments}

GI thanks V.-P. Backlund for useful discussions in the model analysis. RAB is grateful to personnel in the Department of Biomedical Engineering and Computational Science in Aalto University for hospitality in the visits when most of this work was done.

\bibliography{references}

\begin{thebibliography}{10}
\providecommand{\url}[1]{\texttt{#1}}
\providecommand{\urlprefix}{URL }
\expandafter\ifx\csname urlstyle\endcsname\relax
  \providecommand{\doi}[1]{doi:\discretionary{}{}{}#1}\else
  \providecommand{\doi}{doi:\discretionary{}{}{}\begingroup
  \urlstyle{rm}\Url}\fi
\providecommand{\bibAnnoteFile}[1]{%
  \IfFileExists{#1}{\begin{quotation}\noindent\textsc{Key:} #1\\
  \textsc{Annotation:}\ \input{#1}\end{quotation}}{}}
\providecommand{\bibAnnote}[2]{%
  \begin{quotation}\noindent\textsc{Key:} #1\\
  \textsc{Annotation:}\ #2\end{quotation}}
\providecommand{\eprint}[2][]{\url{#2}}

\bibitem{newman2006}
Newman MEJ, Barab{\'a}si AL, Watts DJ (2006) The Structure and Dynamics of
  Networks.
\newblock Princeton: Princeton University Press.
\bibAnnoteFile{newman2006}

\bibitem{caldarelli2007}
Caldarelli G (2007) Scale-Free Networks: Complex Webs in Nature and Technology.
\newblock Oxford: Oxford University Press.
\bibAnnoteFile{caldarelli2007}

\bibitem{dorogovtsev2010}
Dorogovtsev SN (2010) Lectures on Complex Networks.
\newblock Oxford: Oxford University Press.
\bibAnnoteFile{dorogovtsev2010}

\bibitem{castellano2009}
Castellano C, Fortunato S, Loreto V (2009) Statistical physics of social
  dynamics.
\newblock Rev Mod Phys 81: 591--646.
\bibAnnoteFile{castellano2009}

\bibitem{sobkowicz2009}
Sobkowicz P (2009) Modelling opinion formation with physics tools: call for
  closer link with reality.
\newblock J Artif Soc Soc Simul 12: 11.
\bibAnnoteFile{sobkowicz2009}

\bibitem{weidlich1991}
Weidlich W (1991) Physics and social science - the approach of synergetics.
\newblock Phys Rep 204: 1--163.
\bibAnnoteFile{weidlich1991}

\bibitem{holley1975}
Holley RA, Liggett TM (1975) Ergodic theorems for weakly interacting infinite
  systems and the voter model.
\newblock Ann Probab 3: 643--63.
\bibAnnoteFile{holley1975}

\bibitem{sznajd2000}
Sznajd-Weron K, Sznajd J (2000) Opinion evolution in closed community.
\newblock Int J Mod Phys C 11: 1157--65.
\bibAnnoteFile{sznajd2000}

\bibitem{deffuant2000}
Deffuant G, Neau D, Amblard F, Weisbuch G (2000) Mixing beliefs among
  interacting agents.
\newblock Adv Complex Syst 3: 87--98.
\bibAnnoteFile{deffuant2000}

\bibitem{gross2009}
Gross T, Sayama H (2009) Adaptive Networks.
\newblock Heidelberg: Springer Berlin.
\bibAnnoteFile{gross2009}

\bibitem{perc2010}
Perc M, Szolnoki A (2010) Coevolutionary games - a mini review.
\newblock BioSystems 99: 109--25.
\bibAnnoteFile{perc2010}

\bibitem{holme2006}
Holme P, Newman MEJ (2006) Nonequilibrium phase transition in the coevolution
  of networks and opinions.
\newblock Phys Rev E 74: 056108.
\bibAnnoteFile{holme2006}

\bibitem{nardini2008}
Nardini C, Kozma B, Barrat A (2008) Who's talking first? consensus or lack
  thereof in coevolving opinion formation models.
\newblock Phys Rev Lett 100: 158701.
\bibAnnoteFile{nardini2008}

\bibitem{vazquez2008}
Vazquez F, Egu{\'\i}luz VM, San~Miguel M (2008) Generic absorbing transition in
  coevolution dynamics.
\newblock Phys Rev Lett 100: 108702.
\bibAnnoteFile{vazquez2008}

\bibitem{holyst2001sim}
Ho\l{y}st JA, Kacperski K, Schweitzer F (2001) Social impact models of opinion
  dynamics.
\newblock Annu Rev Comput Phys 9: 253--73.
\bibAnnoteFile{holyst2001sim}

\bibitem{kuperman2002srm}
Kuperman M, Zanette D (2002) Stochastic resonance in a model of opinion
  formation on small-world networks.
\newblock Eur Phys J B 26: 387--91.
\bibAnnoteFile{kuperman2002srm}

\bibitem{sobkowicz2009sos}
Sobkowicz P (2009) Studies of opinion stability for small dynamics networks
  with opportunistic agents.
\newblock Int J Mod Phys C 20: 1645--62.
\bibAnnoteFile{sobkowicz2009sos}

\bibitem{iniguez2009}
I{\~n}iguez G, Kert{\'e}sz J, Kaski KK, Barrio RA (2009) Opinion and community
  formation in coevolving networks.
\newblock Phys Rev E 80: 066119.
\bibAnnoteFile{iniguez2009}

\bibitem{kossinets2006}
Kossinets G, Watts DJ (2006) Empirical analysis of an evolving social network.
\newblock Science 311: 88--90.
\bibAnnoteFile{kossinets2006}

\bibitem{iniguez2011cpc}
I{\~n}iguez G, Barrio RA, Kert{\'e}sz J, Kaski KK (2011) Modelling opinion
  formation driven communities in social networks.
\newblock Comput Phys Commun 182: 1866--69.
\bibAnnoteFile{iniguez2011cpc}

\bibitem{oecd2012}
OECD Programme for International Student Assessment. Available at
  http://www.pisa.oecd.org/. Accessed 2012 Feb 8.
\bibAnnoteFile{oecd2012}

\bibitem{miller2004}
Miller JD (2004) Public understanding of, and attitudes toward scientific
  research: what we know and what we need to know.
\newblock Public Understanding of Science 13: 273--94.
\bibAnnoteFile{miller2004}

\bibitem{nisbet2007}
Nisbet MC, Goidel RK (2007) Understanding citizen perceptions of science
  controversy: bridging the ethnography-survey research divide.
\newblock Public Understanding of Science 16: 421--40.
\bibAnnoteFile{nisbet2007}

\bibitem{allum2008}
Allum N, Sturgis P, Tabourazi D, Brunton-Smith I (2008) Science knowledge and
  attitudes across cultures: a meta-analysis.
\newblock Public Understanding of Science 17: 35--54.
\bibAnnoteFile{allum2008}

\bibitem{bennet2011}
Bennett DJ, Jennings RC (2011) Successful Science Communication: Telling it
  like it is.
\newblock Cambridge: Cambridge University Press.
\bibAnnoteFile{bennet2011}

\bibitem{schiele2012}
Schiele B, Claessens M, Shi S (2012) Science Communication in the World:
  Practices, Theories and Trends.
\newblock Heidelberg: Springer Berlin.
\bibAnnoteFile{schiele2012}

\bibitem{cacciatore2011}
Cacciatore MA, Scheufele DA, Corley EA (2011) From enabling technology to
  applications: the evolution of risk perception about nanotechnology.
\newblock Public Understanding of Science 20: 385--404.
\bibAnnoteFile{cacciatore2011}

\bibitem{fiske2010}
Fiske ST, Taylor SE (2010) Social Cognition: From Brains to Culture.
\newblock Singapore: McGraw Hill.
\bibAnnoteFile{fiske2010}

\bibitem{baumeister2011}
Baumeister RF, Bushman BJ (2011) Social Psychology and Human Nature.
\newblock Canada: Wadsworth Cengage Learning.
\bibAnnoteFile{baumeister2011}

\bibitem{himmeli2012}
Himmeli website. Available at http://www.finndiane.fi/software/himmeli/.
  Accessed 2012 Feb 8.
\bibAnnoteFile{himmeli2012}

\bibitem{park2001}
Park HJ (2001) The creation-evolution debate: carving creationism in the public
  media.
\newblock Public Understanding of Science 10: 173--86.
\bibAnnoteFile{park2001}

\bibitem{iniguez2011pre}
I{\~n}iguez G, Kert{\'e}sz J, Kaski KK, Barrio RA (2011) Phase change in an
  opinion-dynamics model with separation of time scales.
\newblock Phys Rev E 83: 016111.
\bibAnnoteFile{iniguez2011pre}

\bibitem{euro2005}
Special Eurobarometer 224 / Wave 63.1 - TNS Opinion \& Social. Available at
  http://ec.europa.eu/public\_opinion/archives/ebs/ebs\_224\_report\_en.pdf.
  Accessed 2012 Feb 8.
\bibAnnoteFile{euro2005}

\bibitem{roper2012}
Roper Center Public Opinion Archives website. Available at
  http://www.ropercenter.uconn.edu/. Accessed 2012 Feb 8.
\bibAnnoteFile{roper2012}

\bibitem{bauer2007}
Bauer MW, Allum N, Miller S (2007) What we can learn from 25 years of survey
  research? liberating and expanding the agents.
\newblock Public Understanding of Science 16: 79--95.
\bibAnnoteFile{bauer2007}

\bibitem{euro2010}
Special Eurobarometer 340 / Wave 73.1 - TNS Opinion \& Social. Available at
  http://ec.europa.eu/public\_opinion/archives/ebs/ebs\_340\_en.pdf. Accessed
  2012 Feb 8.
\bibAnnoteFile{euro2010}

\bibitem{mex2010}
Informe General del Estado de la Ciencia y la Tecnolog{\'\i}a, M{\'e}xico 2010.
  Available at
  http://www.siicyt.gob.mx/siicyt/docs/Estadisticas3/Informe2010/INFORME\_2010%
.pdf. Accessed 2012 Feb 8.
\bibAnnoteFile{mex2010}

\bibitem{bauer2005}
Bauer M (2005) Distinguishing red and green biotechnology: cultivation effects
  of the elite press.
\newblock Int J Public Opin Res 17: 63--89.
\bibAnnoteFile{bauer2005}

\end{thebibliography}


\begin{table}[!ht]
\caption{
\bf{Survey data}}
\begin{tabular}{*{5}{|l}|}
\hline
& \multicolumn{2}{|l|}{EU survey} & \multicolumn{2}{|l|}{Mx survey} \\ \hline
statement & yes (\%) & no (\%) & yes (\%) & no (\%) \\ \hline
a & 21 & 56 & 33 & 63 \\ \hline
b & 22 & 57 & 21 & 76 \\ \hline
c & 24 & 54 & 15 & 81 \\ \hline
d & 35 & 44 & 70 & 17 \\ \hline
e & 38 & 34 & 16 & 84 \\ \hline
f & 40 & 35 & 34 & 64 \\ \hline
g & 44 & 37 & 57 & 41 \\ \hline
h & 46 & 20 & 69 & 23 \\ \hline
i & 53 & 24 & 57 & 39 \\ \hline
j & 58 & 22 & 83 & 14 \\ \hline
k & 61 & 14 & 81 & 15 \\ \hline
l & 66 & 10 & 94 & 3 \\ \hline
m & 72 & 9  & 88 & 9 \\ \hline
n & 75 & 8  & 88 & 9 \\ \hline
o & 78 & 7  & 91 & 6 \\
\hline
\end{tabular}
\begin{flushleft}Statements from the EU and Mx surveys that have been considered for this study. The statement labels are ordered from blatant fallacies to obvious facts, according to our judgment. The percentages of agreement (yes) and disagreement (no) for each statement and survey are also shown.
\end{flushleft}
\label{tab:1}
\end{table}

\end{document}